\documentclass[aps,preprint]{revtex4}
\usepackage{amssymb,epsf}
\usepackage{amsmath}
\usepackage{graphicx}

\begin{document}

\title{Spacetimes with Longitudinal and Angular Magnetic Fields
in Third Order Lovelock Gravity}
\author{M. H. Dehghani$^{1,2}$\footnote{email address:
mhd@shirazu.ac.ir} N. Bostani$^{1}$}
\affiliation{$^1$Physics Department and Biruni Observatory,
College of Sciences, Shiraz
University, Shiraz 71454, Iran\\
$^2$Research Institute for Astrophysics and Astronomy of Maragha
(RIAAM), Maragha, Iran}
\begin{abstract}
We obtain two new classes of magnetic brane solutions in third order Lovelock gravity. The first
class of solutions yields an $(n+1)$-dimensional spacetime with a
longitudinal magnetic field generated by a static source. We
generalize this class of solutions to the case of spinning
magnetic branes with one or more rotation parameters. These
solutions have no curvature singularity and no horizons, but have
a conic geometry. For the spinning brane, when one or more
rotation parameters are nonzero, the brane has a net electric
charge which is proportional to the magnitude of the rotation
parameters, while the static brane has no net electric charge. The
second class of solutions yields a spacetime with an angular
magnetic field. These solutions have no curvature singularity, no
horizon, and no conical singularity. Although the second class of
solutions may be made electrically charged by a boost
transformation, the transformed solutions do not present new
spacetimes. Finally, we use the counterterm method in third order
Lovelock gravity and compute the conserved quantities of these
spacetimes.
\end{abstract}

\maketitle
\section{Introduction}

In recent years asymptotically (anti)-de Sitter [(A)dS] spacetimes
have attracted a great deal of attentions. While the interest on
the asymptotically de Sitter spacetimes comes from the fact that
at the present epoch the Universe expands with acceleration, the
attention to asymptotically AdS spacetimes is due to the fact that
there is a correspondence between supergravity (the low-energy
limit of string theory) in $(n+1)$-dimensional asymptotically AdS
spacetimes and conformal field theory\ (CFT) living on an
$n$-dimensional boundary, known as the AdS/CFT correspondence
\cite{Mal}. The simplest way of having an asymptotically (A)dS
spacetime is to add a cosmological constant term to the right hand
side of Einstein equation. However, the cosmological constant
meets its well known cosmological, fine tuning and coincidence
problems \cite{Cal}. In the context of classical theory of
gravity, the second way of having an asymptotically (A)dS
spacetime is to add higher curvature terms to the left hand side
of Einstein equation. The way that we deal with the asymptotically
(A)dS spacetime is the latter one. Indeed, it seems natural to
reconsider the left hand side of Einstein equation, if one intends
to investigate classical gravity in higher dimensions. The most
natural extension of general relativity in higher dimensional
spacetimes with the assumption of Einstein -- that the left hand
side of the field equations is the most general symmetric
conserved tensor containing no more than two derivatives of the
metric -- is Lovelock theory. Lovelock \cite{Lov} found the most
general symmetric conserved tensor satisfying this property. The
resultant tensor is nonlinear in the Riemann tensor and differs
from the Einstein tensor only if the spacetime has more than 4
dimensions. Since the Lovelock tensor contains metric derivatives
no higher than second order, the quantization of the linearized
Lovelock theory is ghost-free \cite{Zw}

The gravitational action satisfying the assumption of Einstein is
precisely of the form proposed by Lovelock \cite{Lov}:
\begin{equation}
I_{G}=\int d^{d}x\sqrt{-g}\sum_{k=0}^{[d/2]}\alpha _{k}\mathcal{L}_{k},
\label{Lov1}
\end{equation}
where $[z]$ denotes integer part of $z$, $\alpha _{k}$ is an arbitrary
constant and $\mathcal{L}_{k}$ is the Euler density of a $2k$-dimensional
manifold,
\begin{equation}
\mathcal{L}_{k}=\frac{1}{2^{k}}\delta _{\rho _{1}\sigma _{1}\cdots \rho
_{k}\sigma _{k}}^{\mu _{1}\nu _{1}\cdots \mu _{k}\nu _{k}}R_{\mu _{1}\nu
_{1}}^{\phantom{\mu_1\nu_1}{\rho_1\sigma_1}}\cdots R_{\mu _{k}\nu _{k}}^{%
\phantom{\mu_k \nu_k}{\rho_k \sigma_k}}  \label{Lov2}
\end{equation}
In Eq. (\ref{Lov2}) $\delta _{\rho _{1}\sigma _{1}\cdots \rho _{k}\sigma
_{k}}^{\mu _{1}\nu _{1}\cdots \mu _{k}\nu _{k}}$ is the generalized totally
anti-symmetric Kronecker delta and $R_{\mu \nu }^{\phantom{\mu\nu}{\rho%
\sigma}}$ is the Riemann tensor. It is worthwhile to mention that
in $d$ dimensions, all terms for which $k>[d/2]$ are identically
equal to zero, and the term $k=d/2$ is a topological term. So,
only terms for which $k<d/2$ are contributing to the field
equations. In this paper we want to restrict ourself to the first
four terms of Lovelock gravity. The first term is the cosmological
term, the second term is the Einstein term, and the third and
fourth terms are the second order Lovelock (Gauss-Bonnet) and
third order Lovelock terms, respectively. From a geometric point
of view, the combination of these terms in seven-dimensional
spacetimes, is the most general Lagrangian producing second order
field equations, as in the four-dimensional gravity which the
Einstein-Hilbert action is the most general Lagrangian producing
second order field equations. Since the Lovelock Lagrangian
appears in the low energy limit of string theory, there has in
recent years been a renewed interest in Lovelock gravity. In
particular, exact static spherically symmetric black hole
solutions of the Gauss-Bonnet gravity (quadratic in the Riemann
tensor) have been found in Ref. \cite{Des}, and of the
Maxwell-Gauss-Bonnet and Born-Infeld-Gauss-Bonnet models in Ref.
\cite{Wil1}. The thermodynamics of the uncharged static
spherically black hole solutions has been considered in \cite{MS},
of solutions with nontrivial topology in \cite{Cai, Ish} and of
charged solutions in \cite{Wil1,Od1}. Recently NUT charged black
hole solutions of Gauss-Bonnet gravity and Gauss-Bonnet-Maxwell
gravity were obtained \cite{DM}. Not long ago one of us introduced
two new classes of rotating solutions of second order Lovelock
gravity and investigated their thermodynamics \cite{Deh1}, and
made the first attempt for finding exact static
solutions in third order Lovelock gravity with the quartic terms \cite{Deh2}%
, and presented the charged rotating black brane solutions of third order
Lovelock gravity \cite{DM2}.

In this paper we are dealing with the issue of the spacetimes
generated by spinning brane sources in $(n+1)$-dimensional third
order Lovelock theory that are horizonless and have nontrivial
external solutions. These kinds of solutions have been
investigated by many authors in four dimensions. Static uncharged
cylindrically symmetric solutions of Einstein gravity in four
dimensions were considered in \cite{Levi}. Similar static
solutions in the context of cosmic string theory were found in
\cite{Vil}. All of these solutions \cite{Levi,Vil} are horizonless
and have a conical geometry; they are everywhere flat except at
the location of the line source. The extension to include the
electromagnetic field has also been done \cite{Muk,Lem2}. Some
solutions of type IIB supergravity compactified on a four
dimensional torus were considered in \cite{Lun}, which have no
curvature singularity and no conic singularity.

The outline of our paper is as follows. We give a brief review of the field equations
of third order Lovelock gravity in Sec. \ref{Fiel}. In Sec.
\ref{Long} we first present a new class of static horizonless
solutions which produce longitudinal magnetic field, and then
generalize these solutions to the case of spacetimes with one or
more rotation parameters. In Sec. \ref{Angmag} we introduce those
horizonless solutions that produce an angular magnetic field.
Section \ref{Conserv} will be devoted to the use of the
counterterm method to compute the conserved quantities of these
spacetimes. We also compute the electric charge densities of the
branes with rotation or boost parameters. We finish our paper with
some concluding remarks.

\section{Field equations\label{Fiel}}

The main fundamental assumptions in standard general relativity are the
requirements of general covariance and that the field equations for the
metric be second order. Based on the same principles, the Lovelock
Lagrangian is the most general Lagrangian in classical gravity which
produces second order field equations for the metric. The action of third
order Lovelock gravity in the presence of electromagnetic field may be
written as
\begin{eqnarray}
I &=&\int d^{n+1}x\sqrt{-g}\left( -2\Lambda +\mathcal{L}_{1}+\alpha _{2}%
\mathcal{L}_{2}+\alpha _{3}\mathcal{L}_{3}-F^{2}\right)   \nonumber \\
&&+\frac{1}{8\pi }\int_{\partial \mathcal{M}}d^{n}x\sqrt{-\gamma }\left\{
L_{1b}+\alpha _{2}L_{2b}+\alpha _{3}L_{3b}\right\} ,  \label{Act1}
\end{eqnarray}
where $\Lambda $ is the cosmological constant, $\alpha _{2}$ and
$\alpha _{3} $ are Gauss-Bonnet and third order lovelock
coefficients, $\mathcal{L}_{1}=R$ is just the Einstein-Hilbert
Lagrangian, $\mathcal{L}_{2}=R_{\mu \nu \gamma \delta }R^{\mu \nu
\gamma \delta }-4R_{\mu \nu }R^{\mu \nu }+R^{2}$ is the
Gauss-Bonnet Lagrangian and

\begin{eqnarray}
\mathcal{L}_{3} &=&2R^{\mu \nu \sigma \kappa }R_{\sigma \kappa \rho \tau }R_{%
\phantom{\rho \tau }{\mu \nu }}^{\rho \tau }+8R_{\phantom{\mu \nu}{\sigma
\rho}}^{\mu \nu }R_{\phantom {\sigma \kappa} {\nu \tau}}^{\sigma \kappa }R_{%
\phantom{\rho \tau}{ \mu \kappa}}^{\rho \tau }+24R^{\mu \nu \sigma \kappa
}R_{\sigma \kappa \nu \rho }R_{\phantom{\rho}{\mu}}^{\rho }  \nonumber \\
&&+3RR^{\mu \nu \sigma \kappa }R_{\sigma \kappa \mu \nu }+24R^{\mu \nu
\sigma \kappa }R_{\sigma \mu }R_{\kappa \nu }+16R^{\mu \nu }R_{\nu \sigma
}R_{\phantom{\sigma}{\mu}}^{\sigma }-12RR^{\mu \nu }R_{\mu \nu }+R^{3}
\label{L3}
\end{eqnarray}
is the third order Lovelock Lagrangian. In Eq. (\ref{Act1}) $F^{2}=F_{\mu
\nu }F^{\mu \nu }$ where $F_{\mu \nu }=\partial _{\mu }A_{\nu }-\partial
_{\nu }A_{\mu }$ is electromagnetic tensor field and $A_{\mu }$ is the
vector potential. The second integral in Eq. (\ref{Act1}) is a boundary term
which is chosen such that the variational principle is well defined \cite
{DM2, Myers, DBSH}. In this integral $L_{1b}=K$, $L_{2b}=2(J-2\widehat{G}%
_{ab}^{(1)}K^{ab})$ and
\begin{eqnarray*}
L_{3b} &=&3(P-2\widehat{G}_{ab}^{(2)}K^{ab}-12\widehat{R}_{ab}J^{ab}+2%
\widehat{R}J \\
&&-4K\widehat{R}_{abcd}K^{ac}K^{bd}-8\widehat{R}%
_{abcd}K^{ac}K_{e}^{b}K^{ed}),
\end{eqnarray*}
where $\gamma _{\mu \nu }$ is induced metric on the boundary, $K$ is trace
of extrinsic curvature of boundary, $\widehat{G}_{ab}^{(1)}$ and $\widehat{G}%
_{ab}^{(2)}$ are the $n$-dimensional Einstein and second
order Lovelock tensors of the metric
$\gamma _{ab}$ [$\widehat{G}%
_{ab}^{(2)}$ will be given in Eq. (\ref{Love2})] and $J$ and $P$ are the trace of
\begin{equation}
J_{ab}=\frac{1}{3}%
(2KK_{ac}K_{b}^{c}+K_{cd}K^{cd}K_{ab}-2K_{ac}K^{cd}K_{db}-K^{2}K_{ab})
\end{equation}
and
\begin{eqnarray}
P_{ab} &=&\frac{1}{5}%
\{[K^{4}-6K^{2}K^{cd}K_{cd}+8KK_{cd}K_{e}^{d}K^{ec}-6K_{cd}K^{de}K_{ef}K^{fc}+3(K_{cd}K^{cd})^{2}]K_{ab}
\nonumber \\
&&-(4K^{3}-12KK_{ed}K^{ed}+8K_{de}K_{f}^{e}K^{fd})K_{ac}K_{b}^{c}-24KK_{ac}K^{cd}K_{de}K_{b}^{e}
\nonumber \\
&&+(12K^{2}-12K_{ef}K^{ef})K_{ac}K^{cd}K_{db}+24K_{ac}K^{cd}K_{de}K^{ef}K_{bf}\}
\label{Pab}
\end{eqnarray}

Since in Lovelock gravity, only terms for which $k<d/2$ are contributing to
the field equations and we want to consider the third order lovelock
gravity, therefore we consider the $d$-dimensional spacetimes with $d\geq 7$%
. Varying the action (\ref{Act1}) with respect to the metric tensor $g_{\mu
\nu }$ and electromagnetic tensor field $F_{\mu \nu }$ the equations of
gravitation and electromagnetic fields are obtained as:
\begin{equation}
G_{\mu \nu }^{(1)}+\Lambda g_{\mu \nu }+\alpha _{2}G_{\mu \nu
}^{(2)}+\alpha _{3}G_{\mu \nu }^{(3)}=T_{\mu \nu },  \label{Geq}
\end{equation}
\begin{equation}
\\
\nabla _{\nu }F^{\mu \nu }=0,  \label{EMeq}
\end{equation}
where $T_{\mu \nu }=2F_{\phantom{\lambda}{\mu}}^{\rho }F_{\rho \nu }-\frac{1%
}{2}F_{\rho \sigma }F^{\rho \sigma }g_{\mu \nu }$ is the energy-momentum
tensor of electromagnetic field, $G_{\mu \nu }^{(1)}$ is just the Einstein
tensor, and $G_{\mu \nu }^{(2)}$ and $G_{\mu \nu }^{(3)}$ are the second and
third order Lovelock tensors given as \cite{Hoi}:
\begin{equation}
G_{\mu \nu }^{(2)}=2(R_{\mu \sigma \kappa \tau }R_{\nu }^{\phantom{\nu}%
\sigma \kappa \tau }-2R_{\mu \rho \nu \sigma }R^{\rho \sigma }-2R_{\mu
\sigma }R_{\phantom{\sigma}\nu }^{\sigma }+RR_{\mu \nu })-\frac{1}{2}%
\mathcal{L}_{2}g_{\mu \nu },  \label{Love2}
\end{equation}
\begin{eqnarray}
G_{\mu \nu }^{(3)} &=&-3(4R^{\tau \rho \sigma \kappa }R_{\sigma \kappa
\lambda \rho }R_{\phantom{\lambda }{\nu \tau \mu}}^{\lambda }-8R_{%
\phantom{\tau \rho}{\lambda \sigma}}^{\tau \rho }R_{\phantom{\sigma
\kappa}{\tau \mu}}^{\sigma \kappa }R_{\phantom{\lambda }{\nu \rho \kappa}%
}^{\lambda }+2R_{\nu }^{\phantom{\nu}{\tau \sigma \kappa}}R_{\sigma \kappa
\lambda \rho }R_{\phantom{\lambda \rho}{\tau \mu}}^{\lambda \rho }  \nonumber
\\
&&-R^{\tau \rho \sigma \kappa }R_{\sigma \kappa \tau \rho }R_{\nu \mu }+8R_{%
\phantom{\tau}{\nu \sigma \rho}}^{\tau }R_{\phantom{\sigma \kappa}{\tau \mu}%
}^{\sigma \kappa }R_{\phantom{\rho}\kappa }^{\rho }+8R_{\phantom
{\sigma}{\nu \tau \kappa}}^{\sigma }R_{\phantom {\tau \rho}{\sigma \mu}%
}^{\tau \rho }R_{\phantom{\kappa}{\rho}}^{\kappa }  \nonumber \\
&&+4R_{\nu }^{\phantom{\nu}{\tau \sigma \kappa}}R_{\sigma \kappa \mu \rho
}R_{\phantom{\rho}{\tau}}^{\rho }-4R_{\nu }^{\phantom{\nu}{\tau \sigma
\kappa }}R_{\sigma \kappa \tau \rho }R_{\phantom{\rho}{\mu}}^{\rho
}+4R^{\tau \rho \sigma \kappa }R_{\sigma \kappa \tau \mu }R_{\nu \rho
}+2RR_{\nu }^{\phantom{\nu}{\kappa \tau \rho}}R_{\tau \rho \kappa \mu }
\nonumber \\
&&+8R_{\phantom{\tau}{\nu \mu \rho }}^{\tau }R_{\phantom{\rho}{\sigma}%
}^{\rho }R_{\phantom{\sigma}{\tau}}^{\sigma }-8R_{\phantom{\sigma}{\nu \tau
\rho }}^{\sigma }R_{\phantom{\tau}{\sigma}}^{\tau }R_{\mu }^{\rho }-8R_{%
\phantom{\tau }{\sigma \mu}}^{\tau \rho }R_{\phantom{\sigma}{\tau }}^{\sigma
}R_{\nu \rho }-4RR_{\phantom{\tau}{\nu \mu \rho }}^{\tau }R_{\phantom{\rho}%
\tau }^{\rho }  \nonumber \\
&&+4R^{\tau \rho }R_{\rho \tau }R_{\nu \mu }-8R_{\phantom{\tau}{\nu}}^{\tau
}R_{\tau \rho }R_{\phantom{\rho}{\mu}}^{\rho }+4RR_{\nu \rho }R_{%
\phantom{\rho}{\mu }}^{\rho }-R^{2}R_{\nu \mu })-\frac{1}{2}\mathcal{L}%
_{3}g_{\mu \nu }  \label{Love3}
\end{eqnarray}

\section{Longitudinal magnetic field solutions}\label{Long}
Here, we want to obtain the $(n+1)$-dimensional solutions of Eqs. (\ref{Geq}%
)-(\ref{EMeq}) which produce longitudinal magnetic fields in the Euclidean
submanifold spans by the $x^{i}$ coordinates ($i=1,...,n-2$). We assume that
the metric has the following form:
\begin{equation}
ds^{2}=-\frac{\rho ^{2}}{l^{2}}dt^{2}+\frac{d\rho ^{2}}{f(\rho )}%
+l^{2}f(\rho )d\phi ^{2}+\frac{\rho ^{2}}{l^{2}}dX^{2}{,}  \label{Met1a}
\end{equation}
where $dX^{2}={{\sum_{i=1}^{n-2}}}(dx^{i})^{2}$. Note that the coordinates $%
x^{i}$ have the dimension of length, while the angular coordinate $\phi $ is
dimensionless as usual and ranges in $0\leq \phi <2\pi $. The motivation for
this metric gauge $[g_{tt}\varpropto -\rho ^{2}$ and $(g_{\rho \rho
})^{-1}\varpropto g_{\phi \phi }]$ instead of the usual Schwarzschild gauge $%
[(g_{\rho \rho })^{-1}\varpropto g_{tt}$ and $g_{\phi \phi }\varpropto \rho
^{2}]$ comes from the fact that we are looking for a horizonless solution
with conic singularity.

The gauge potential is given by

\begin{equation}
A_{\mu }=-\frac{2}{(n-2)}\frac{ql^{n-1}}{\rho ^{n-2}}\delta _{\mu }^{\phi }
\label{pot1}
\end{equation}
To find the function $f(\rho )$, one may use any components of Eq. (\ref{Geq}%
). The simplest equation is the $\rho \rho $ component of these equations
which can be written as
\begin{eqnarray}
&&\left[ 180(_{\phantom{n}{5}}^{n-1})\alpha _{3}\rho f^{2}-6(_{\phantom{n}{3}%
}^{n-1})\alpha _{2}f\rho ^{3}+\frac{n-1}{2}\rho ^{5}\right]
f^{\prime
}+\Lambda r^{6}  \nonumber \\
&&+360(_{\phantom{n}{6}}^{n-1})\alpha _{3}f^{3}-12(_{\phantom{n}{4}%
}^{n-1})\alpha _{2}\rho ^{2}f^{2}+(_{\phantom{n}{2}}^{n-1})\rho
^{4}f=4q^{2}l^{2n-4}\rho ^{8-2n},  \label{Eqf}
\end{eqnarray}
where prime denotes the derivative with respect to $\rho $. The
general solution of Eq. (\ref{Eqf}) is
\begin{equation}
f(\rho )=\frac{\rho ^{2}}{b_{3}\alpha }\left\{ \left( \sqrt{\gamma
+k^{2}(\rho )}+k(\rho )\right) ^{1/3}-\gamma ^{1/3}\left( \sqrt{\gamma
+k^{2}(\rho )}+k(\rho )\right) ^{-1/3}+b_{2}\right\} ,  \label{f1}
\end{equation}
where
\begin{equation}
\alpha _{2}=b_{2}\frac{\alpha }{(n-2)(n-3)},\quad \alpha _{3}=b_{3}\frac{%
\alpha ^{2}}{72(_{\phantom{n}{4}}^{n-2})},\qquad \gamma =\
(b_{3}-b_{2}^{2})^{3},
\end{equation}
and
\begin{equation}
k(\rho )=-\frac{1}{2}b_{2}(3b_{3}-2b_{2}^{2})+3\alpha b_{3}^{2}\left( -\frac{%
\Lambda }{n(n-1)}+\frac{m}{\rho
^{n}(n-1)}-\frac{4q^{2}}{(n-1)(n-2)\rho ^{2n-2}l^{4-2n}}\right)
\label{kkk}
\end{equation}
The constants $m$ and $q$ in Eq. (\ref{kkk}) are the mass and charge
parameters of the metric which are related to the mass and charge density of
the solution. The function $k(\rho )$ approaches a constant as $\rho $ goes
to infinity, and the effective cosmological constant for the spacetime
is
\[
\Lambda _{\mathrm{eff}}=-\frac{1}{\alpha b_{3}}\left\{ b_{2}+{\left( \sqrt{%
\gamma +\lambda ^{2}}+\lambda \right) ^{1/3}-}\gamma ^{1/3}\left( \sqrt{%
\gamma +\lambda ^{2}}+\lambda \right) ^{-1/3}\right\} ,
\]
where
\begin{equation}
\lambda =-\frac{1}{2}b_{2}(3b_{3}-2b_{2}^{2})-\frac{3b_{3}^{2}\alpha \Lambda
}{n(n-1)}  \label{lambda}
\end{equation}
In the rest of the paper, we only investigate the case of $\gamma \geq 0$.
In this case $\Lambda _{\mathrm{eff}}$ and $f(\rho )$ are real. The function
$f(\rho )$ is negative for large values of $\rho $, if $\Lambda _{\mathrm{eff%
}}>0$. Since $g_{\rho \rho }$ and $g_{\phi \phi }$ are related by $f(\rho
)=g_{\rho \rho }^{-1}=l^{-2}g_{\phi \phi }$, and therefore when $g_{\rho
\rho }$ becomes negative (which occurs for large $\rho $) so does $g_{\phi
\phi }$. This leads to an apparent change of signature of the metric from $%
(n-1)+$ to $(n-2)+$ as $\rho $ goes to infinity, which is not allowed. Thus,
$\Lambda _{\mathrm{eff}}$ should be negative which occurs provided
\begin{equation}
{\lambda }>{}-\frac{1}{2}b_{2}(3b_{3}-2b_{2}^{2})  \label{lamcon}
\end{equation}
Equation (\ref{lambda}) shows that the condition (\ref{lamcon}) is satisfied
if $\Lambda <0$.

In order to study the general structure of the solution given in Eq. (\ref
{f1}), we first look for curvature singularities. It is easy to show that
the Kretschmann scalar $R_{\mu \nu \lambda \kappa }R^{\mu \nu \lambda \kappa
}$ diverge at $\rho =0$ and therefore one might think that there is a
curvature singularity located at $\rho =0$. However, as we will see below,
the spacetime will never achieve $\rho =0$. The function $f(\rho )$ is
negative for $\rho <r_{+}$ and positive for $\rho >r_{+}$, where $r_{+}$ is
the largest root of $f(\rho )=0$, which may be written as
\[
(n-2)\rho ^{2n-2}l^{4-2n}\Lambda -(n-2)\rho ^{n-2}l^{4-2n}mn+4q^{2}n=0
\]
Again $g_{\rho \rho }$ cannot be negative (which occurs for$\rho <r_{+}$),
because of the change of signature of the metric from $(n-1)+$ to $(n-2)+$.
Thus, one cannot extend the spacetime to $\rho <r_{+}$. To get rid of this
incorrect extension, we introduce the new radial coordinate $r$ as
\[
r^{2}=\rho ^{2}-r_{+}^{2}\Rightarrow d\rho ^{2}=\frac{r^{2}}{r^{2}+r_{+}^{2}}%
dr^{2}
\]
With this new coordinate, the metric (\ref{Met1a}) is
\begin{equation}
ds^{2}=-\frac{r^{2}+r_{+}^{2}}{l^{2}}dt^{2}+\frac{r^{2}}{%
(r^{2}+r_{+}^{2})f(r)}dr^{2}+l^{2}f(r)d\phi ^{2}+\frac{r^{2}+r_{+}^{2}}{l^{2}%
}dX^{2},  \label{Metr1b}
\end{equation}
where the coordinates $r$ and $\phi $ assume the values $0\leq r<\infty $
and $0\leq \phi <2\pi $, and $f(r)$ is now given as
\begin{equation}
f(r)=\frac{(r^{2}+r_{+}^{2})}{b_{3}\alpha }\left\{ \left( \sqrt{\gamma
+k^{2}(r)}+k(r)\right) ^{1/3}-\gamma ^{1/3}\left( \sqrt{\gamma +k^{2}(r)}%
+k(r)\right) ^{-1/3}+b_{2}\right\} ,  \label{FF2}
\end{equation}
where
\begin{equation}
k(r)=-\frac{1}{2}b_{2}(3b_{3}-2b_{2}^{2})+3\alpha b_{3}^{2}\left( -\frac{%
\Lambda }{n(n-1)}+\frac{m}{(r^{2}+r_{+}^{2})^{n/2}(n-1)}-%
\frac{4q^{2}}{(n-1)(n-2)(r^{2}+r_{+}^{2})^{(n-1)}l^{4-2n}}\right)
\label{KKK}
\end{equation}
The gauge potential in the new coordinate is
\begin{equation}
A_{\mu }=-\frac{2}{(n-2)}\frac{ql^{(n-1)}}{(r^{2}+r_{+}^{2})^{(n-2)/2}}%
\delta _{\mu }^{\phi }
\end{equation}
The function $f(r)$ given in Eq. (\ref{FF2}) is positive in the whole
spacetime and is zero at $r=0$. Also note that the Kretschmann scalar does
not diverge in the range $0\leq r<\infty $. Therefore this spacetime has no
curvature singularities and no horizons. However, it has a conic geometry
and has a conical singularity at $r=0$, since:

\[
\lim_{r\rightarrow 0}\frac{1}{r}\sqrt{\frac{g_{\phi \phi }}{g_{rr}}}%
=lr_{+}G_{0},
\]
where

\begin{eqnarray}
G_{0} &=&\frac{r_{+}^{2}}{3b_{3}^{2}\alpha }(\frac{k_{0}^{\prime \prime }}{%
\sqrt{4b_{3}-3b_{2}^{2}}})\left\{ b_{2}+\left( 4[b_{3}\sqrt{4b_{3}-3b_{2}^{2}%
}-b_{2}(3b_{3}-2b_{2}^{2})]\right) ^{1/3}\right\} ,  \notag \\
k_{0}^{\prime \prime } &=&k^{\prime \prime }(r=0)=-3\alpha
b_{3}^{2}\left(
\frac{mn}{(n-1)r_{+}^{n+2}}-\frac{8q^{2}}{(n-2)l^{4-2n}r_{+}^{2n}}\right)
, \label{G0}
\end{eqnarray}

which is not equal to one. That is, as the radius $r$ tends to
zero, the limit of the ratio ``\textrm{circumference/radius}'' is
not $2\pi $ and therefore the spacetime has a conical singularity
at $r=0$.

Of course, one may ask for the completeness of the spacetime with $r\geq 0$
(or $\rho \geq r_{+}$). It is easy to see that the spacetime described by
Eq. (\ref{Metr1b}) is both null and timelike geodesically complete \cite
{Lem2,Hor}. In fact, we can show that every null or timelike geodesic
starting from an arbitrary point can either extend to infinite values of the
affine parameter along the geodesic or end on a singularity at $r=0$. Using
the geodesic equation, one obtains
\begin{eqnarray}
\dot{t} &=&\frac{l^{2}}{r^{2}+r_{+}^{2}}E,\hspace{0.5cm}\dot{x^{i}}=\frac{%
l^{2}}{r^{2}+r_{+}^{2}}P^{i},\hspace{0.5cm}\dot{\phi}=\frac{1}{l^{2}f(r)}L,
\\
r^{2}\dot{r}^{2} &=&(r^{2}+r_{+}^{2})f(r)\left[ \frac{l^{2}(E^{2}-\mathbf{P}%
^{2})}{r^{2}+r_{+}^{2}}-\eta \right] -\frac{r^{2}+r_{+}^{2}}{l^{2}}L^{2},
\label{Geo2}
\end{eqnarray}
where the overdot denotes the derivative with respect to an affine
parameter, and $\eta $ is zero for null geodesics and $+1$ for timelike
geodesics. $E$, $L$, and $P^{i}$'s are the conserved quantities associated
with the coordinates $t$, $\phi $, and $x^{i}$, respectively, and $\mathbf{P}%
^{2}=\sum_{i=1}^{n-2}(P^{i})^{2}$. Notice that $f(r)$ is always positive for
$r>0$ and zero for $r=0$. First we consider the null geodesics ($\eta =0$).
(i) If $E^{2}>\mathbf{P}^{2}$ the spiraling particles ($L>0$) coming from
infinity have a turning point at $r_{tp}>0$, while the nonspiraling
particles ($L=0$) have a turning point at $r_{tp}=0$. (ii) If $E^{2}=\mathbf{%
P}^{2}$ and $L=0$, whatever the value of $r$, $\dot{r}$ and $\dot{\phi}$
vanish and therefore the null particles moves on the $z$-axis. (iii) For $%
E^{2}=\mathbf{P}^{2}$ and $L\neq 0$, and also for $E^{2}<\mathbf{P}^{2}$ and
any values of $L$, there is no possible null geodesic. Second, we analyze
the timelike geodesics ($\eta =+1$). Timelike geodesics is possible only if $%
l^{2}(E^{2}-\mathbf{P}^{2})>r_{+}^{2}$. In this case spiraling ($L\neq 0$)
timelike particles are bound between $r_{tp}^{a}$ and $r_{tp}^{b}$ given by $%
0<r_{tp}^{a}\leq r_{tp}^{b}<\sqrt{l^{2}(E^{2}-\mathbf{P}^{2})-r_{+}^{2}}$,
while the turning points for the nonspiraling particles ($L=0$) are $%
r_{tp}^{1}=0$ and $r_{tp}^{2}=\sqrt{l^{2}(E^{2}-\mathbf{P}^{2})-r_{+}^{2}}$
\bigskip

\subsection{Longitudinal magnetic field solutions with all rotation
parameters\label{Lmag2}}

The rotation group in $n+1$ dimensions is $SO(n)$ and therefore the number
of independent rotation parameters is $[n/2]$, where $[x]$ is the integer
part of $x$. We now generalize the above solution given in Eq. (\ref{Metr1b}%
) with $k\leq \lbrack n/2]$ rotation parameters. This generalized solution
can be written as
\begin{eqnarray}
ds^{2} &=&-\frac{r^{2}+r_{+}^{2}}{l^{2}}\left( \Xi dt-{{\sum_{i=1}^{k}}}%
a_{i}d\phi ^{i}\right) ^{2}+f(r)\left( \sqrt{\Xi ^{2}-1}dt-\frac{\Xi }{\sqrt{%
\Xi ^{2}-1}}{{\sum_{i=1}^{k}}}a_{i}d\phi ^{i}\right) ^{2}  \nonumber \\
&&+\frac{r^{2}dr^{2}}{(r^{2}+r_{+}^{2})f(r)}+\frac{r^{2}+r_{+}^{2}}{%
l^{2}(\Xi ^{2}-1)}{\sum_{i<j}^{k}}(a_{i}d\phi _{j}-a_{j}d\phi _{i})^{2}+%
\frac{r^{2}+r_{+}^{2}}{l^{2}}dX^{2},  \label{Metr2}
\end{eqnarray}
where $\Xi =\sqrt{1+\sum_{i}^{k}a_{i}^{2}/l^{2}}$, $dX^{2}$ is the Euclidean
metric on the $(n-k-1)$-dimensional submanifold and $f(r)$ is the same as $%
f(r)$ given in Eq. (\ref{FF2}). The gauge potential is
\begin{equation}
A_{\mu }=\frac{2}{(n-2)}\frac{ql^{(n-2)}}{(r^{2}+r_{+}^{2})^{(n-2)/2}}\left(
\sqrt{\Xi ^{2}-1}\delta _{\mu }^{0}-\frac{\Xi }{\sqrt{\Xi ^{2}-1}}%
a_{i}\delta _{\mu }^{i}\right) ;\hspace{0.5cm}{\text{(no sum on i)}}
\label{Pot2}
\end{equation}
Again this spacetime has no horizon and curvature singularity. However, it
has a conical singularity at $r=0$.

\section{Angular magnetic field solutions\label{Angmag}}

In Sec. \ref{Long} we found a spacetime generated by a magnetic
source which produces a longitudinal magnetic field along $x^{i}$
coordinates. In this section we want to obtain a spacetime
generated by a magnetic source that produce angular magnetic
fields along the $\phi ^{i}$ coordinates. Following the steps of
Sec. \ref{Long} but now with the roles of $\phi $ and $x$
interchanged, we can directly write the metric and vector
potential satisfying the field equations (\ref{Geq})-(\ref{EMeq})
as
\begin{eqnarray}
ds^{2} &=&-\frac{r^{2}+r_{+}^{2}}{l^{2}}dt^{2}+\frac{r^{2}dr^{2}}{%
(r^{2}+r_{+}^{2})f(r)}  \nonumber \\
&&+(r^{2}+r_{+}^{2}){{\sum_{i=1}^{n-2}}}(d\phi
^{i})^{2}+f(r)dx^{2},\label{met4}
\end{eqnarray}
where $f(r)$ is given in Eq. (\ref{FF2}). The angular coordinates $\phi ^{i}$%
's range in $0\leq \phi ^{i}<2\pi $. The gauge potential is now given by
\begin{equation}
A_{\mu }=-\frac{2}{(n-2)}\frac{ql^{(n-2)}}{(r^{2}+r_{+}^{2})^{(n-2)/2}}%
\delta _{\mu }^{x}
\end{equation}
The Kretschmann scalar does not diverge for any $r$ and therefore
there is no curvature singularity. The spacetime (\ref{met4}) is
also free of conic singularity. In addition, it is notable to
mention that the radial geodesic passes through $r=0$ (which is
free of singularity) from positive values to negative values of
the coordinate $r$. This shows that the radial coordinate in Eq.
(\ref{met4}) can take the values $-\infty <r<\infty $. This
analysis may suggest that one is in the presence of a traversable
wormhole with a
throat of dimension $r_{+}$. However, in the vicinity of $r=0$, the metric (%
\ref{met4}) can be written as

\begin{eqnarray}
ds^{2} &=&-\frac{r_{+}^{2}}{l^{2}}dt^{2}+\frac{1}{r_{+}^{2}}%
G_{0}^{-1}dr^{2}+r_{+}^{2}{{\sum_{i=1}^{n-2}}}(d\phi ^{i})^{2}  \nonumber \\
&&+G_{0}r^{2}dx^{2},
\end{eqnarray}
where $G_{0}$ is given in Eq. (\ref{G0}). This clearly shows that, at $r=0$,
the $x$ direction collapses and therefore we have to abandon the wormhole
interpretation.

To add linear momentum to the spacetime along the coordinate $x^{i}$, we
perform the boost transformation
\[
t\mapsto \Xi t-(v_{i}/l)x^{i};\text{ \ \ }x^{i}\mapsto \Xi x^{i}-(v_{i}/l)t%
\text{ \ \ (no sum on }i\text{)}
\]
in the $t-x_{i}$ plane, where $v_{i}\ $is a boost parameter and $\Xi =\sqrt{%
1+\sum_{i}^{\kappa }v_{i}^{2}/l^{2}}$ ($i$ can run from $1$ to $\kappa \leq
n-2$). One obtains
\begin{eqnarray}
ds^{2} &=&-\frac{r^{2}+r_{+}^{2}}{l^{2}}R^{2}(r)\left( \Xi dt-l^{-1}{{%
\sum_{i=1}^{\kappa }}}v_{i}dx^{i}\right) ^{2}+f(r)\left( \sqrt{\Xi ^{2}-1}dt-%
\frac{\Xi }{l\sqrt{\Xi ^{2}-1}}{{\sum_{i=1}^{\kappa }}}v_{i}dx^{i}\right)
^{2}  \nonumber \\
&&+\frac{r^{2}+r_{+}^{2}}{l^{4}(\Xi ^{2}-1)}R^{2}(r)\text{ }{%
\sum_{i<j}^{\kappa }}(v_{i}dx_{j}-v_{j}dx_{i})^{2}+\frac{r^{2}dr^{2}}{%
(r^{2}+r_{+}^{2})f(r)}+(r^{2}+r_{+}^{2})R^{2}(r)d\Omega
^{2}\label{met5}
\end{eqnarray}
The gauge potential is given by

\begin{equation}
A_{\mu }=\frac{qb^{(3-n)\gamma }}{\Gamma (r^{2}+r_{+}^{2})^{\Gamma /2}}%
\left( \sqrt{\Xi ^{2}-1}\delta _{\mu }^{t}-\frac{\Xi }{l\sqrt{\Xi ^{2}-1}}%
v_{i}\delta _{\mu }^{i}\right) \hspace{0.5cm}{\text{(no sum on i)}}
\end{equation}
This boost
transformation is permitted globally since $x^{i}$ is not an
angular coordinate. Thus the boosted solution (\ref{met5}) is not
a new solution. However, it generates an electric field.

\section{Conserved Quantities\label{Conserv}}

In general, the action (\ref{Act1}) is divergent when evaluated on
solutions, as is the Hamiltonian and other associated conserved
quantities. In Einstein gravity, one can remove the non
logarithmic divergent terms in the action by adding a counterterm
action $I_{\mathrm{ct}} $ which is a functional of the boundary
curvature invariants \cite{Kraus}. The issue of determination of
boundary counterterms with their coefficients for higher-order
Lovelock theories is at this point an open question.
However for the case of a boundary with zero curvature [$\widehat{R}%
_{abcd}(\gamma )=0$], it is quite straightforward. This is because all
curvature invariants are zero except for a constant, and so the only
possible boundary counterterm is one proportional to the volume of the
boundary regardless of the number of dimensions \cite{DM2,DBSH}:
\begin{equation}
I_{\mathrm{ct}}=\frac{1}{8\pi }\int_{\partial \mathcal{M}}d^{n}x\sqrt{%
-\gamma }\frac{n-1}{L},  \label{metc}
\end{equation}
where $L$ is a scale length factor that depends on $l$, $\alpha _{2}$ and $%
\alpha _{3}$, that must reduce to $l$ as $\alpha _{2}$ and $\alpha _{3}$ go
to zero. Having the total finite action, one can use the quasilocal
definition \cite{BY,BCM} to construct a divergence free stress-energy
tensor. For the case of manifolds with zero curvature boundary the finite
stress energy tensor is
\begin{eqnarray}
T^{ab} &=&\frac{1}{8\pi }\{(K^{ab}-K\gamma ^{ab})+2\alpha
_{2}(3J^{ab}-J\gamma ^{ab})  \nonumber \\
&&\ +3\alpha _{3}(5P^{ab}-P\gamma ^{ab})+\frac{n-1}{L}\gamma ^{ab}\ \}
\label{Stres}
\end{eqnarray}
The first three terms in Eq. (\ref{Stres}) result from the
variation of the surface terms in action (\ref{Act1}) with respect
to $\gamma ^{ab}$, and the last term is the counterterm that is
the variation of $I_{ct}$ with respect to $\gamma ^{ab}$. To
compute the conserved charges of the
spacetime, we choose a spacelike surface $\mathcal{B}$ in $\partial \mathcal{%
M}$ with metric $\sigma _{ij}$, and write the boundary metric in ADM form:
\begin{equation}
\gamma _{ab}dx^{a}dx^{a}=-N^{2}dt^{2}+\sigma _{ij}\left( d\phi
^{i}+V^{i}dt\right) \left( d\phi ^{j}+V^{j}dt\right) ,
\end{equation}
where the coordinates $\phi ^{i}$ are the angular variables
parameterizing the hypersurface of constant $r$ around the origin,
and $N$ and $V^{i}$ are the lapse and shift functions
respectively. When there is a Killing vector field $\mathcal{\xi
}$ on the boundary, then the quasilocal conserved quantities
associated with the stress tensors of Eq. (\ref{Stres}) can be
written as
\begin{equation}
\mathcal{Q}(\mathcal{\xi )}=\int_{\mathcal{B}}d^{n-1}\varphi \sqrt{\sigma }%
T_{ab}n^{a}\mathcal{\xi }^{b},  \label{charge}
\end{equation}
where $\sigma $ is the determinant of the metric $\sigma _{ij}$,
and $n^{a}$ is the timelike unit normal vector to the boundary
$\mathcal{B}${.} In the context of counterterm method, the limit
in which the boundary $\mathcal{B}$ becomes infinite
($\mathcal{B}_{\infty }$) is taken, and the counterterm
prescription ensures that the action and conserved charges are
finite. No embedding of the surface $\mathcal{B}$ in to a
reference of spacetime is required and the quantities which are
computed are intrinsic to the spacetimes.

For our case, the magnetic solutions of third order Lovelock gravity, the first Killing
vector is $\xi =\partial /\partial t$, and therefore its associated
conserved charge is the total mass. Denoting the volume of the hypersurface
boundary $\mathcal{B}$ at constant $t$ and $r$ by $V_{n-1}$ the mass per
unit volume $V_{n-1}$ is
\begin{equation}
M=\int_{\mathcal{B}}d^{n-1}x\sqrt{\sigma }T_{ab}n^{a}\xi ^{b}=\frac{1}{8\pi }%
ml^{p-1}\left[ n(\Xi ^{2}-1)+1\right] ,
\end{equation}
where $p$ is the number of angular coordinates $\phi ^{i}$ of the spacetime.
For the case of spacetimes with a longitudinal magnetic field, the charges
associated with the rotational Killing symmetries generated by $\zeta
_{i}=\partial /\partial \phi ^{i}$ are the components of angular
momentum per unit volume $V_{n-1}$ of the system calculated as
\begin{equation}
J_{i}=\int_{\mathcal{B}}d^{n-1}x\sqrt{\sigma }T_{ab}n^{a}\zeta _{i}^{b}=%
\frac{1}{8\pi }n\Xi l^{p-1}ma_{i}  \label{Ang}
\end{equation}
In the case of the spacetimes with an angular magnetic field
introduced in Sec. \ref{Angmag}, we encounter conserved quantities
associated with translational Killing symmetries generated by
$\varsigma _{i}=\partial /\partial x^{i}$. These conserved
quantities are the components of linear momentum per unit volume $V_{n-1}$ computed as

\begin{equation}
P_{i}=\int_{\mathcal{B}}d^{n-1}x\sqrt{\sigma }T_{ab}n^{a}\varsigma _{i}^{b}=%
\frac{1}{8\pi }n\Xi l^{p-2}mv_{i}
\end{equation}

Next, we calculate the electric charge of the solutions. To determine the
electric field we should consider the projections of the electromagnetic
field tensor on special hypersurfaces. The normal to such hypersurfaces for
the spacetimes with a longitudinal magnetic field is
\[
u^{0}=\frac{1}{N}, \hspace{.5cm} u^{r}=0,\hspace{.5cm} u^{i}=-\frac{N^{i}}{N},
\]
and the electric field is $E^{\mu }=g^{\mu \rho }F_{\rho \nu }u^{\nu }$.
Then the electric charge per unit volume $V_{n-1}$ can be found by calculating the flux of the
electromagnetic field at infinity, yielding
\begin{equation}
Q=\frac{V_{n-1}}{4\pi }ql^{p-1}\sqrt{\Xi ^{2}-1}  \label{elecch}
\end{equation}
Note that the electric charge is proportional to the magnitude of
rotation parameters or boost parameters, and is zero for the case
of a static solution.

\section{Closing Remarks}

In this paper, we added the second and third order Lovelock terms
to the Einstein-Maxwell action with a negative cosmological
constant. We introduced two classes of solutions which are
asymptotically anti-de Sitter. The first class of solutions yields
an $(n+1)$-dimensional spacetime with a longitudinal magnetic
field [the only nonzero component of the vector potential is
$A_{\phi }(r)$] generated by a static magnetic brane. We also
generalized these solutions to the case of rotating spacetimes with a
longitudinal magnetic field. We found that these solutions have no
curvature singularity and no horizons, but have conic singularity
at $r=0$. In these spacetimes, when all the rotation parameters
are zero (static case), the electric field vanishes, and therefore
the brane has no net electric charge. For the spinning brane, when
one or more rotation parameters are nonzero, the brane has a net
electric charge density which is proportional to the magnitude of
the rotation parameter given by $\sqrt{\Xi ^{2}-1}$. The second
class of solutions yields a spacetime with angular magnetic field.
These solutions have no curvature singularity, no horizon, and no
conic singularity. Again, we found that the branes in these
spacetimes have no net electric charge when all the boost
parameters are zero. We also showed that, for the case of
traveling branes with nonzero boost parameters, the net electric
charge density of the brane is
proportional to the magnitude of the velocities of the brane ($\sqrt{\Xi ^{2}-1%
}$).

The counterterm method inspired by the AdS/CFT correspondence conjecture has
been widely applied to the case of Einstein gravity. Here we applied this
method to the case of third order Lovelock gravity and calculated the
conserved quantities of the two classes of solutions. We found that the
counterterm has only one term, since the boundaries of our spacetimes are
curvature-free. Other related problems such as the application of the
counterterm method to the case of solutions of higher curvature gravity with
nonzero curvature boundary remain to be carried out.

\begin{center}
\textbf{ACKNOWLEDGMENTS}
\end{center}

This work has been supported by Research Institute for Astrophysics and
Astronomy of Maragha, Iran.

\end{document}